\documentclass{elsart}
\usepackage{graphicx}
\def \be {\begin{equation}}
\def \ee {\end{equation}}
 
\begin{document}

\begin{frontmatter}
\title{Why only few are so successful ?}
\author{P. K. Mohanty}
\address{TCMP Division, Saha Institute of Nuclear Physics \\1/AF Bidhan Nagar, Kolkata 700064 India}
\ead{pk.mohanty@saha.ac.in}
  
\begin{abstract}
In many professons employees  are rewarded  according to 
their relative performance. Corresponding economy can be 
modeled by taking $N$ independent 
agents who gain from the market with a rate which depends on their current 
gain. We argue that this simple realistic rate  generates a scale free 
distribution even though intrinsic ability of agents are marginally 
different from each other. As an evidence we provide distribution of 
scores for two different systems (a) the global stock game where players invest in real 
stock market and (b) the international cricket.  
\end{abstract}

\end{frontmatter}

\section{Introduction}
\label{intro}

In equilibrium systems power law distributions are observed at  the criticality. 
Many open and driven systems in nature, however,  naturally self organize to produce 
scale free distributions\cite{Bak}. Distribution of rain fall, magnitude of earthquake, 
link distribution of world wide web  are few examples to mention. 
The scale invariant distributions are also seen in social and economic 
systems\cite{econo}. About a century ago Vilferdo Pareto pointed out 
that wealth $w$ in any society is distributes as 
$P(w)= w^{-\gamma}$. Several attempts have been made earlier to understand 
this phenomena, namely 'the Pareto-law for the wealthy people'\cite{Pareto}. 
An interesting analogy  has been drawn \cite{Yako00} between the economic system 
and the system of ideal gases where particles and their energies are 
modelled as agent and their wealth, and redistribution of energy during collision is 
modelled as trading between agents.  This analogy, which naturally generate  
Gibb's distribution, could successfully  explain 97\% of the observed  income 
distribution. The rich (about 3\%), however, follow a scale-free distribution 
which was explained later \cite{CCM}  using ideal-gas like models.  

   In this article we argue that economy works differently at different levels. 
In particular for rich and  successful it is quite a different game. There are  
certain kind of  occupations, for example the law,  the medicine and the 
journalism, market pays individuals not according to their 
absolute performance, but according to their performance relative to 
others in the same occupation.  The same is true in the sports, share and 
entertainment industries. These systems where {\it "winner plays an important 
role in  the market"} may be named as celebrity markets (CM)\cite{celebrity}.

  How does a system (or market)  generates a successful professional or celebrity ? 
Of course, there are exceptionally brilliant and strategic individuals who play and 
controls the market. But often, strategy of players or agents in the market are 
not very different from each other. However,  the distribution of their 
success or wealth is highly asymmetric with power law tails, where most become 
unsuccessful and only a few  become successful. To  understand this phenomena
we introduce a  simple model of $N$ agents in section \ref {themodel}. In section 
\ref{evidence} the theoretical results  are compared  in certain example  systems 
belongs to this class, namely celebrity markets. Finally  the conclusion and 
some discussions are given in section \ref{conclusion}.

\section{The model}
\label{themodel}
  Let us take an unbiased sample of $N$ agents, labeled by $i=1,2,\dots,N$, 
who invests equal amount in the market ({\it say}, stock market). The  net gain 
of the agents$\{m_i\}$  are  taken  to be integers for simplicity  and 
set  $\{m_i=0\}$ at time $t=0$.  In each time step $dt$ a randomly chosen 
agent $j$ first decides with probability $z$, if he wants  to continue 
investing. With probability $1-z$ the agent become inactive forever. If active, 
the net gain of the agent $m_j$  is increased by unity, with a probability 
$w(m_j)$. Of course,  $\{ w(m_j)\}$ are normalized such that total 
probability of all active agents is unity.

It is reasonable to assume that the growth 
rate of agents $w(m)$ depends on the the instantaneous gain $m$ and that 
$w(m)$ is an increasing function. Because, if $m_j(t) > m_i(t)$, agent $j$ 
can be considered strategically smarter (who studies the market better) 
than agent $i$ at time $t$ and thus $w(m_j) > w(m_i)$. 

Note that, there is no direct interection between agents. The only interaction 
comes from the fact that the growth rate is relative. Thus our model is an 
ensemble of $N$ independent agents where wealth $m$ of an agent follows 
a discrete time dynamics  
\begin{equation}
(m-1)  \mathop{\longrightarrow}^{zw(m)} m,
\end{equation}
where $w(m)$ is an increasing function.  
Depending on their asymptotic limits, increasing functions 
may be  classified  into two categories;  
(a) when $w(\infty)$ is $\infty$ and (b) $w(\infty)$ is  finite (say, unity). 
It is obvious that for case (a), the growth rate $w(m)$  for  
the smartest agent, chosen stochastically by the process, 
is large compared to the rest and thus he predictively  wins the market. 
In the second case, where $w(m)$ is a marginally 
increasing function, probability of gaining an extra  unit is 
comparable among agents, which mimics the cometition existing in real 
markets. Moreover, we have assumed  that the agents are strategically similar 
and thus choice (b) is more appropriate.

Since agents are independent,
$Prob.(\{m_i\}) = \prod_i p(m_i)$, where $p(m)$ is the 
probability that an agent gains $m$ unit of wealth.  
To gain $m$ units, one must go through the process  
$0\to 1, 1\to 2,  \dots (m-1) \to m$, which occurs with rate 
$w(1), w(2)  \dots w(m)$ respectively. Thus 
the normalized probability is, 
\begin{equation}
p(m)=  \frac{z^m\prod_{k=1}^{m} w(k)}{ F(z)}~~\textrm{where }~~ F(z)=  p(0) + \sum_{m=1} z^m\prod_{k=1}^{m}w(k). \label{eq:gce}
\end{equation}

The average gain   $\rho(z)=\langle m \rangle =zF^\prime(z)/F(z)$
is monotonically increases starting from $\rho(0) =0$. 
Since maximum value of $z$ is $1$ (when agents   
keep on investing indefinitely), the maximum average gain is $\rho(1)$. 
If $\rho(1)=\infty$, one can fix any arbitrary density by suitably 
choosing $z$. But when $\rho(1)$ is finite, say $\rho(1)=\rho_c$, it is 
impossible to have uniform macroscopic density $\rho>\rho_c$. 
Thus in this case, the extra gain $(\rho-\rho_c)N$ would be owned
only by  one or few agents.  In next section we will 
discuss about such a  possibility, namely the  condensation of wealth.

\subsection{Condensation ?}
Let us take $z=1$. Then agents  do not have  a choice but to  invest  
indefinitely. Let us further impose an condition that the agents 
keep on investing until total gain becomes $\sum m_i=M$. 
Now, the partition function  of an ensemble of systems with  total gain 
$M$ being conserved is then \be Q_M =\prod_{k=1}^{M} w(k).\label{eq:ce} \ee 
One may consider (\ref{eq:ce}) as a canonical partition function and 
clearly  (\ref{eq:gce}) represents the grand canonical partition 
function of this system with fugacity $z$. 

In this  canonical ensemble one can choose the density $\rho=M/N$ 
arbitrarily large. When $\rho>\rho_c$, with 
$\rho_c$ being finite, we have extra wealth $(\rho-\rho_c)N$ which 
can not be distributed macroscopically. Some agent(s) would gain this macroscopic 
amount. It can be argued\cite{zrpmap} that in the thermodynamic limit, wealth would 
preferably go to one agent instead being distributed between few agents.

The possibility of having condensation (or a {\it  super celebrity}) depends on  
the rate $w(k)$. First we need that $\rho_c= \lim_{z\to 1} \frac{zF^\prime(z)}{F(z)}$ 
is finite. Since $F(z)$ is analytic for $z<1$, we must check whether 
$zF^\prime(z)= \sum_1^\infty mz^m Q_m$ is finite as $z\to 1$. This series  will  
converge, when ratio of successive terms decay more slowly than $1+1/m$. 
Thus asymptotically $w(m)$ should increase faster  than $1-2/m$. 
It is evident from the  Taylor's series of  $w(m) =   1- w_1/m - w_2/m^2 \dots$ that 
condensation would occur when $w_1>2$. 

To demonstrate the condensation, let us make a simple 
choice \be w(m) = m/(m+b)\label{eq:w}\ee which is a marginally increasing 
function. In this case $w_1=b$ and thus condensation occurs 
for large density $\rho>\rho_c$, if $b>2$. To calculate $\rho_c$, 
first note that  
\be Q_m= \frac{\Gamma(m+1)\Gamma(b+1)}{\Gamma(m+b+1)}\label{eq:Q}\ee 
Thus, $F(1) =  b/(b-1)$ and $F^\prime(1)= F(1)/(b-2)$ 
and hence, $\rho_c= F^\prime(1)/F(1)= (b-2)^{-1}$.

Any other choice of rate  where coefficient of $m^{-1}$ in Taylor's 
series of $w(m)$ is $-b$ is similar to (\ref{eq:w}) except that  
$\rho_c$ is different from $(b-2)^{-1}$.
Thus we will continue further discussions with choice (\ref{eq:w}) .

Distribution of wealth can be obtained from   
(\ref{eq:Q}). Asymptotically, $Q(m) \propto m^{-b}$. Thus, up to 
a normalization constant $F(z)^{-1}$,  
\be
p(m) = z^m Q(m) = z^m m^{-b}.
\label{eq:pareto}
\ee 
Note that $p(m)$ is similar to observed economic distribution: 
an exponential distribution  for small $m$ and a power-law 
in the tail. 
 
From (\ref{eq:pareto}) one can 
show that $\langle m\rangle$ diverges for $b>2$ and thus, in this case 
condensation occurs for sufficiently large densities. Since, 
many observed distributions follow  Pareto law $p(m) \propto m^{-b}$ 
with $2<b<3$, condenseation of wealth is expected in these economic 
systems if per capita income (i. e., the density $\rho$) is very large. 
Condensation in economic systems has been observed and  modelled
earlier\cite{wealth-cond}. It was argued that the macroscopic 
accumulation may be viewed as existing  corruptions in societies. 
Here, we show that  such  a phenomena can occur naturally in CMs.

\section{Evidence}
\label{evidence}
  As we have discussed earlier, in many occupations, like the medicine 
the journalism, the share trading and in the entertainment industries like 
the sports and the film  industries, {\it "winner takes all the market".} 
In this section we would cite some examples  which are close to the model 
discussed here. 
%

\subsection{Global stock game}

\begin{figure}
\begin{center}
\includegraphics*[width=8cm]{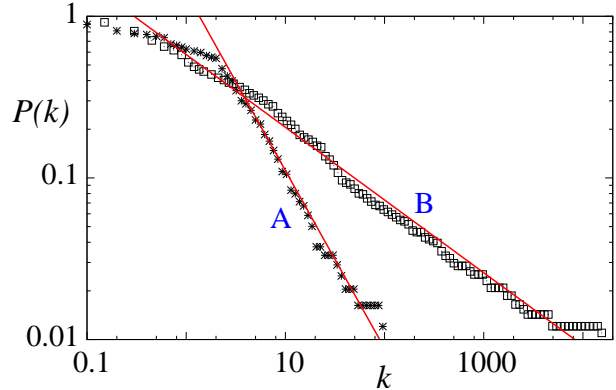}
\end{center}
\caption{Probability of gaining more than $k\%$ in global stock game\cite{gsg}: 
(A)  Group of K-12 students (total$373$ members, started in January, 2006, 
starting wealth is \$10,000 per student). (B) Group of $1294$ members, 
started in January 2005, starting wealth is \$666666666 per person.}
\label{fig:gsg}
\end{figure}

  Let us take an hypothetical  example where brokers invest the  
same amount of money in stock market. What would 
be the distribution of their net gain ? It is difficult to carry out 
such a controlled experiment where (real) money is involved. 
However a prototype experiment is an existing game, namely global 
stock game (gsg)\cite{gsg}. When  a group  joins this game  they  get  
a fixed amount of {\it gsg  dollars} to make transactions in the real  
stock market (NSE) and individual  NAV is evaluated. If these 
money would have been real the player could have earned the NAV. 
This game is usually played by  (A) a group of  school/college students 
as an learing exercise or (B) by a group of brokers to get experienced  
in thereal stock market without loosing money.  We have collected data 
for both the groups. 

\textit{Group A :}  We have combined two group of K-12 students  
and  calculated the percentage of gain $m$. If probability of net 
gain $p(m)\sim m^{-b}$, the probability of gaining more than $k\%$
is $P(k) \sim k^{-b+1}$. Thus, from the plot of  $P(k)$ 
versus $k$  in log scale one can obtain $b$. Condensation is expected 
for $b>2$. In this case,  $P(k)$ do not vanish for large $k$ 
and  a correct fitting function is  $P(k) = c_1+ c_2 k^{-b+1}$. 
In fact for  group A  (see Fig. \ref{fig:gsg}) we find $b=2.10\pm .05$, 
with $c_1=0.06$ and $c_2=1.35$. 

\textit{Group B :} This a single group of $1294$ members joined  to win the 
contest {\it money666}.  For this group we find (see Fig. \ref{fig:gsg})
that  $b=1.43 \pm .05$, with $c_1=0$ and $c_2=0.52$. 
 
Note  the exponent $b$ is different for different groups. A possible 
reason for group $A$ to have $b>2$ is that it consists of beginners, 
where some learners are smarter than  others  and  play well enough to 
become the super celebrity. However group B consists   of exparts (probably) 
and thus nobody gains  exceptionally different from others, which 
explains the smaller value of $b$ 

\subsection{International cricket}
 Another example is  the score (e. g., run) of cricketers in 
international cricket, a bat and ball sport 
played between two teams of eleven players each. 
Readers unfamiliar with the game   
may look at {\it http://en.wikipedia.org/wiki/Cricket} for 
details. In this example, let us take two cricketers who started 
playing about the same time. Obviously one who scores better gets 
selected for the next international match. Thus, the rate of 
increase of score $w(k)$ is an increasing function. 
Again, the difference between 
rates of two cricketers already having huge runs is small and hence 
the choice  (\ref{eq:w})is  reasonable. 

\begin{figure}
\begin{center}
\includegraphics[width=6cm]{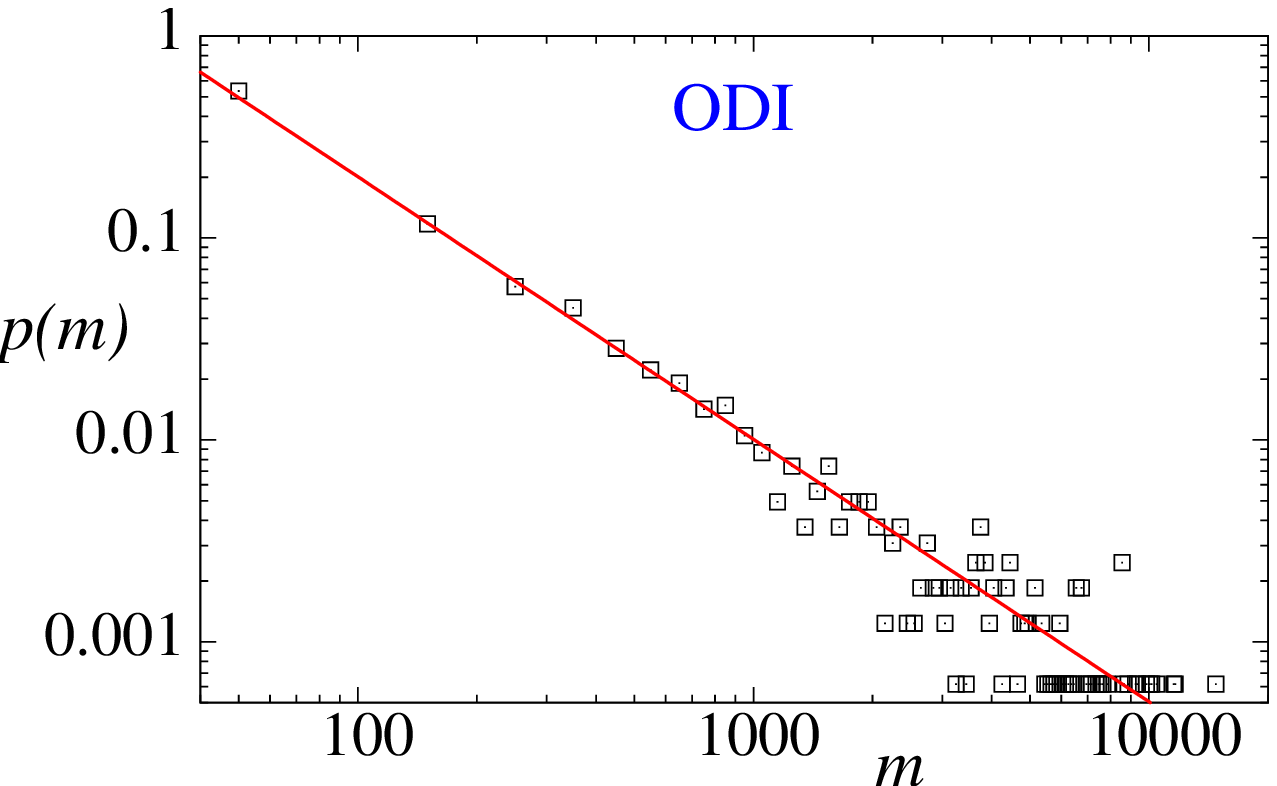} 
\includegraphics[width=6cm]{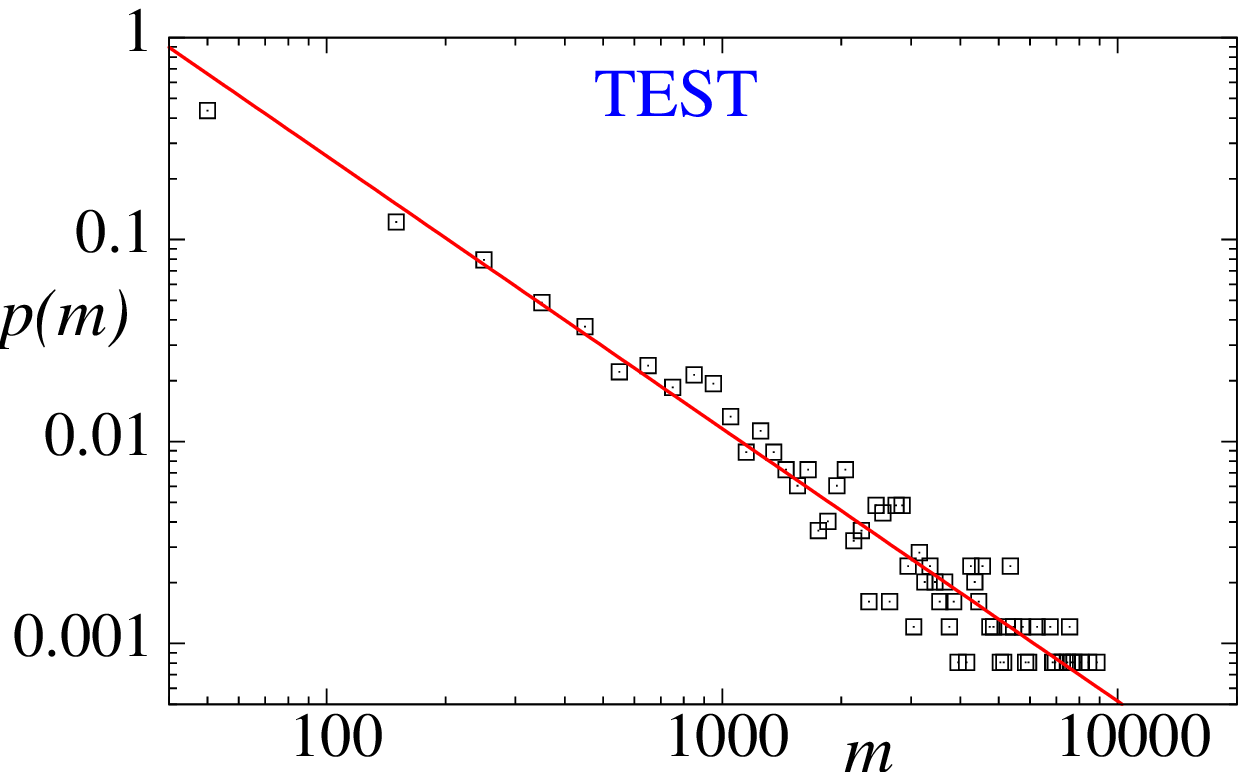}
\end{center}
\caption{Distribution of runs: (ODI) The one day international cricket, played by 
$21$ countries having $1620$ players;  (TEST) The  test cricket which is played by 
$10$ countries with  $2482$ players. Data is collected from 
{\it http://www.howstat.com}. }
\label{fig:cricket}
\end{figure}

We have collected life time score (run) of the international
cricketers  for both one day international (ODI) 
 and the test   cricket.   
The calculated  distribution of runs $p(m)$ are  plotted  
in log scale (Fig. \ref{fig:cricket}). Corresponding slopes are found 
to be $b=1.3\pm .05$ for ODI  and $b=1.35\pm .05$ for test.
Since $b<2$, we do not expect a super celebrity here. It means, 
we will never  fine an exceptional cricketer (both in  ODI and in 
test cricket) who could score strikingly different from others.
 
\section{Conclusion and Discussion}
\label{conclusion}
In this paper we have introduced a model a markets, where 
agents (employees or players) are rewarded according to their comparative 
performance. A player who is successful at present has a better 
probability of being successful in future. We show that 
this underlying mechanism "success comes easily to people who are 
already successful"  generates a skew distribution even though  the 
ability of  players are marginally different from each other. The 
model successfully describes observation of wealth condensation in
economic systems. It also predicts that, in cricket, a brilliant 
performance which is strikingly different from others, is not 
expected from any player.

 This model is general enough to describe different systems having 
different $b$. It would be nice to  obtain  data at different 
intermediate times so that from the evolution of the net gain one 
can calculate $w(k)$ and thus $b$ directly. Such studies would 
certainly justify the model  better. There are many other systems 
which are test ground for CM model. One  example is the distribution 
of citation of different papers (not authors)  where a better cited
article is expected to get more citations in future. Another example 
is the distribution of chromosomal changes per tumor\cite{cancer} in 
different kinds of cancers which show a power-law. Here, the mechanism 
is the following. A cell having more aberrant chromosomes $m$,  
would generate daughter cells with more aberrations compared to a 
cell having smaller $m$. Thus (\ref{eq:w}) is a reasonable choice.

  It is worth noting that the  CM model can be  mapped to a well-known  
model in non-equilibrium  
studies namely zero range process (ZRP)\cite{ZRP}. 
ZRP  is defined  on a one dimensional periodic lattice with 
$N$ sites and   $M$ 
particles initially distributed randomly among sites. The dynamics of 
the model is as follows. One particle is transfered from a randomly chosen 
site to it's  rightward neighbour with a rate $u(m)$ where $m$ is the number
of particles  in the departure site.  It can be shown, that the steady 
state distribution of particles follow (\ref{eq:Q}) in canonical  or 
(\ref{eq:gce}) in grand canonical ensemble with  $w(m) = u(m)^{-1}.$
The condensation criteria of  CM model is identical to that 
of the ZRP. 

   The celebrity market model introduced here  can also be used to model 
anomalous diffusion, which will be discussed elsewhere.


\begin{thebibliography}{99}
\bibitem{Bak}    \textit{How Nature Works: The science of self-organized criticality}, by Per Bak (Springer-Verlag, New York, 1996).
\bibitem {econo} Econophysics of Wealth Distribution, 
edited by A. chatterjee, S. Yarlagadda, and B. K. Chakraborti 
(Springer Verlag, Milan, 2005).
\bibitem{Pareto} V. Pareto, Cours d'economie Politique (F. Rouge, lausanne, 1897).
\bibitem{Yako00} A. A. Dragulescu and V. M. Yakovenko, Physica {\bf A 299}, 213 (2001); {\it ibid},   Eur. Phys. J. {\bf B 17}, 723 (2000).
\bibitem{CCM} A. Chakraborti and B. K. Chakrabarti, Eur. Phys. J. {\bf B 17}, 
167 (2000); P. K. Mohanty,  Phys. Rev. E {\bf 74}, 011117 (2006).
\bibitem{celebrity}  Here, a successful professional is  named as celebrity, 
independent of whether he is renowned in the society or not.
\bibitem{zrpmap}  Equation (\ref{eq:gce}) is identical to the steady state 
distribution of particles in zero range process. Criteria for 
condensation  and few other arguments presented here  can also be found in the 
review article\cite{ZRP}.  
\bibitem{wealth-cond}J.-P. Bouchard, and   M. Mezard,  Physica A {\bf 282}, 536 (2000);
Z. Burda, D. Johnston, J. Jurkiewicz, M. Kami\'nski, M. A. Nowak, G. Papp, and I. Zahed, 
Phys. Rev. E {\bf 65}, 026102 (2002).
\bibitem{gsg} Data is collected from {\it http://www.stocksquest.com}.
\bibitem{cancer} A. Frigyesi, D. Gisselsson, F. Mitelman, and M. H\"oglund, Cancer Research \textbf{63}, 7094 (2003).
\bibitem{ZRP} For a recent review see, M. R. Evans, T. Hanney, J. Phys. A: Math. Gen. 
{\bf 38}  R195(2005).

\end{thebibliography}
\end{document}